\documentclass[aps,pr,twocolumn,superscriptaddress,showpacs,preprintnumbers,amsmath,amssymb,fleqn]{revtex4}

\usepackage{graphicx} 
\usepackage{dcolumn}  

\graphicspath{{ps}}

\begin{document}






\title{ \quad\\[1.0cm] Measurements of the absolute branching fractions of $B^{+} \to X_{c\bar{c}} K^{+}$ and $B^{+} \to  \bar{D}^{(\ast) 0} \pi^{+} $ at Belle}

\noaffiliation
\affiliation{University of the Basque Country UPV/EHU, 48080 Bilbao}
\affiliation{Beihang University, Beijing 100191}
\affiliation{Budker Institute of Nuclear Physics SB RAS, Novosibirsk 630090}
\affiliation{Faculty of Mathematics and Physics, Charles University, 121 16 Prague}
\affiliation{Chonnam National University, Kwangju 660-701}
\affiliation{University of Cincinnati, Cincinnati, Ohio 45221}
\affiliation{Deutsches Elektronen--Synchrotron, 22607 Hamburg}
\affiliation{University of Florida, Gainesville, Florida 32611}
\affiliation{Justus-Liebig-Universit\"at Gie\ss{}en, 35392 Gie\ss{}en}
\affiliation{Gifu University, Gifu 501-1193}
\affiliation{SOKENDAI (The Graduate University for Advanced Studies), Hayama 240-0193}
\affiliation{Gyeongsang National University, Chinju 660-701}
\affiliation{Hanyang University, Seoul 133-791}
\affiliation{University of Hawaii, Honolulu, Hawaii 96822}
\affiliation{High Energy Accelerator Research Organization (KEK), Tsukuba 305-0801}
\affiliation{J-PARC Branch, KEK Theory Center, High Energy Accelerator Research Organization (KEK), Tsukuba 305-0801}
\affiliation{IKERBASQUE, Basque Foundation for Science, 48013 Bilbao}
\affiliation{Indian Institute of Science Education and Research Mohali, SAS Nagar, 140306}
\affiliation{Indian Institute of Technology Bhubaneswar, Satya Nagar 751007}
\affiliation{Indian Institute of Technology Guwahati, Assam 781039}
\affiliation{Indian Institute of Technology Madras, Chennai 600036}
\affiliation{Indiana University, Bloomington, Indiana 47408}
\affiliation{Institute of High Energy Physics, Chinese Academy of Sciences, Beijing 100049}
\affiliation{Institute of High Energy Physics, Vienna 1050}
\affiliation{Institute for High Energy Physics, Protvino 142281}
\affiliation{University of Mississippi, University, Mississippi 38677}
\affiliation{INFN - Sezione di Napoli, 80126 Napoli}
\affiliation{INFN - Sezione di Torino, 10125 Torino}
\affiliation{Advanced Science Research Center, Japan Atomic Energy Agency, Naka 319-1195}
\affiliation{J. Stefan Institute, 1000 Ljubljana}
\affiliation{Kanagawa University, Yokohama 221-8686}
\affiliation{Institut f\"ur Experimentelle Kernphysik, Karlsruher Institut f\"ur Technologie, 76131 Karlsruhe}
\affiliation{Kennesaw State University, Kennesaw, Georgia 30144}
\affiliation{King Abdulaziz City for Science and Technology, Riyadh 11442}
\affiliation{Department of Physics, Faculty of Science, King Abdulaziz University, Jeddah 21589}
\affiliation{Korea Institute of Science and Technology Information, Daejeon 305-806}
\affiliation{Korea University, Seoul 136-713}
\affiliation{Kyoto University, Kyoto 606-8502}
\affiliation{Kyungpook National University, Daegu 702-701}
\affiliation{\'Ecole Polytechnique F\'ed\'erale de Lausanne (EPFL), Lausanne 1015}
\affiliation{P.N. Lebedev Physical Institute of the Russian Academy of Sciences, Moscow 119991}
\affiliation{Faculty of Mathematics and Physics, University of Ljubljana, 1000 Ljubljana}
\affiliation{Ludwig Maximilians University, 80539 Munich}
\affiliation{Luther College, Decorah, Iowa 52101}
\affiliation{University of Malaya, 50603 Kuala Lumpur}
\affiliation{University of Maribor, 2000 Maribor}
\affiliation{Max-Planck-Institut f\"ur Physik, 80805 M\"unchen}
\affiliation{School of Physics, University of Melbourne, Victoria 3010}
\affiliation{University of Miyazaki, Miyazaki 889-2192}
\affiliation{Moscow Physical Engineering Institute, Moscow 115409}
\affiliation{Moscow Institute of Physics and Technology, Moscow Region 141700}
\affiliation{Graduate School of Science, Nagoya University, Nagoya 464-8602}
\affiliation{Kobayashi-Maskawa Institute, Nagoya University, Nagoya 464-8602}
\affiliation{Nara Women's University, Nara 630-8506}
\affiliation{National Central University, Chung-li 32054}
\affiliation{National United University, Miao Li 36003}
\affiliation{Department of Physics, National Taiwan University, Taipei 10617}
\affiliation{H. Niewodniczanski Institute of Nuclear Physics, Krakow 31-342}
\affiliation{Nippon Dental University, Niigata 951-8580}
\affiliation{Niigata University, Niigata 950-2181}
\affiliation{Novosibirsk State University, Novosibirsk 630090}
\affiliation{Osaka City University, Osaka 558-8585}
\affiliation{Pacific Northwest National Laboratory, Richland, Washington 99352}
\affiliation{University of Pittsburgh, Pittsburgh, Pennsylvania 15260}
\affiliation{Theoretical Research Division, Nishina Center, RIKEN, Saitama 351-0198}
\affiliation{University of Science and Technology of China, Hefei 230026}
\affiliation{Showa Pharmaceutical University, Tokyo 194-8543}
\affiliation{Soongsil University, Seoul 156-743}
\affiliation{Stefan Meyer Institute for Subatomic Physics, Vienna 1090}
\affiliation{Sungkyunkwan University, Suwon 440-746}
\affiliation{School of Physics, University of Sydney, New South Wales 2006}
\affiliation{Department of Physics, Faculty of Science, University of Tabuk, Tabuk 71451}
\affiliation{Tata Institute of Fundamental Research, Mumbai 400005}
\affiliation{Excellence Cluster Universe, Technische Universit\"at M\"unchen, 85748 Garching}
\affiliation{Department of Physics, Technische Universit\"at M\"unchen, 85748 Garching}
\affiliation{Toho University, Funabashi 274-8510}
\affiliation{Department of Physics, Tohoku University, Sendai 980-8578}
\affiliation{Earthquake Research Institute, University of Tokyo, Tokyo 113-0032}
\affiliation{Department of Physics, University of Tokyo, Tokyo 113-0033}
\affiliation{Tokyo Institute of Technology, Tokyo 152-8550}
\affiliation{Tokyo Metropolitan University, Tokyo 192-0397}
\affiliation{University of Torino, 10124 Torino}
\affiliation{Virginia Polytechnic Institute and State University, Blacksburg, Virginia 24061}
\affiliation{Wayne State University, Detroit, Michigan 48202}
\affiliation{Yamagata University, Yamagata 990-8560}
\affiliation{Yonsei University, Seoul 120-749}
  \author{Y.~Kato}\affiliation{Kobayashi-Maskawa Institute, Nagoya University, Nagoya 464-8602} 
  \author{T.~Iijima}\affiliation{Kobayashi-Maskawa Institute, Nagoya University, Nagoya 464-8602}\affiliation{Graduate School of Science, Nagoya University, Nagoya 464-8602} 
  \author{I.~Adachi}\affiliation{High Energy Accelerator Research Organization (KEK), Tsukuba 305-0801}\affiliation{SOKENDAI (The Graduate University for Advanced Studies), Hayama 240-0193} 
  \author{H.~Aihara}\affiliation{Department of Physics, University of Tokyo, Tokyo 113-0033} 
  \author{S.~Al~Said}\affiliation{Department of Physics, Faculty of Science, University of Tabuk, Tabuk 71451}\affiliation{Department of Physics, Faculty of Science, King Abdulaziz University, Jeddah 21589} 
  \author{D.~M.~Asner}\affiliation{Pacific Northwest National Laboratory, Richland, Washington 99352} 
  \author{V.~Aulchenko}\affiliation{Budker Institute of Nuclear Physics SB RAS, Novosibirsk 630090}\affiliation{Novosibirsk State University, Novosibirsk 630090} 
  \author{T.~Aushev}\affiliation{Moscow Institute of Physics and Technology, Moscow Region 141700} 
  \author{R.~Ayad}\affiliation{Department of Physics, Faculty of Science, University of Tabuk, Tabuk 71451} 
  \author{V.~Babu}\affiliation{Tata Institute of Fundamental Research, Mumbai 400005} 
% ...
\collaboration{The Belle Collaboration}

\begin{abstract}
We present the measurement of the absolute branching fractions of $B^{+} \to X_{c\bar{c}}  K^{+}$ and $B^{+} \to \bar{D}^{(\ast) 0}  \pi^{+} $ decays, using a  data sample of $772\times10^{6}$ $B\bar{B}$ pairs collected at the $\Upsilon(4S)$ resonance with the
Belle detector at the KEKB asymmetric-energy $e^{+}e^{-}$ collider.
Here, $X_{c\bar{c}}$ denotes $\eta_{c}$, $J/\psi$, $\chi_{c0}$, $\chi_{c1}$, $\eta_{c}(2S)$, $\psi(2S)$, $\psi(3770)$, $X(3872)$, and $X(3915)$.
We do not observe significant signals for $X(3872)$ nor $X(3915)$, and set the 90$\%$ confidence level upper limits:
${\cal B}(B^{+} \to X(3872)  K^{+} )<2.6 \times 10^{-4}$ and ${\cal B}(B^{+} \to X(3915)  K^{+} )<2.8 \times 10^{-4}$.
These represent the most stringent upper limit for ${\cal B}(B^{+} \to X(3872)  K^{+} )$ to date and the first limit for ${\cal B}(B^{+} \to X(3915)  K^{+} )$.
The measured branching fractions for $\eta_{c}$ and $\eta_{c}(2S)$ are the most precise to date:
${\cal B}(B^{+} \to \eta_{c} K^{+} )=(12.0\pm0.8\pm0.7) \times 10^{-4}$ and ${\cal B}(B^{+} \to \eta_{c}(2S)K^{+})  =(4.8\pm1.1\pm0.3) \times 10^{-4}$
, where the first and second uncertainties are statistical and systematic, respectively.
\end{abstract}


\maketitle

\tighten

{\renewcommand{\thefootnote}{\fnsymbol{footnote}}}
\setcounter{footnote}{0}

\section{Introduction}\label{section_intro}
The discovery of the $X(3872)$ by the Belle collaboration \cite{Choi:2003ue} opened a new era in the field of hadron spectroscopy.
The $X(3872)$ does not correspond to any of the predicted charmonium states in the quark model~\cite{Olsen:2004fp}, and is an exotic hadron candidate.
The most natural interpretation of $X(3872)$ is a $D^{0}$ and $\bar{D}^{\ast 0}$ molecular state,
as its mass is quite close to the combined mass of these charmed mesons and 
it has $J^{PC}=1^{++}$ \cite{Aaij:2013zoa}, which is consistent 
with an S-wave $D^{0} \bar{D}^{\ast 0}$ molecule state. However, a pure molecular interpretation cannot explain the large cross section 
observed by the CDF experiment in $p\bar{p}$ collisions at $\sqrt{s}=1.9$ TeV \cite{Abulencia:2006ma}.
Therefore, the most plausible explanation of $X(3872)$ is an admixture of a molecular state
and a pure charmonium $\chi_{c1}(2P)$ \cite{Takizawa:2012hy}.
The large value of the ratio  $ {\cal B}(X(3872) \to \psi(2S) \gamma)/ {\cal B}(X(3872) \to J/\psi \gamma)$
 \cite{Aaij:2014ala,Aubert:2008ae} 
and the lack of observation of $\chi_{c1}(2P) \to \chi_{c1} \pi^{+} \pi^{-}$~\cite{Bhardwaj:2015rju} also support this interpretation.
In order to understand the nature of $X(3872)$, a measurement of the absolute branching fraction  ${\cal B}(B^{+} \to X(3872) K^{+} )$
is quite useful \cite{Zanetti:2011ju}. With this measurement in hand, we can determine 
${\cal B} (X(3872) \to f)$, where $f$ is a possible final state, since the product of 
${\cal B}(B^{+} \to X(3872) K^{+} )$ and ${\cal B} (X(3872) \to f)$ is measured~\cite{PDG}.
The measurement of ${\cal B}(B^{+} \to X(3872) K^{+} )$ is possible in $B$-factory experiments operating at a center-of-mass energy that corresponds to 
the mass of the $\Upsilon(4S)$ resonance, where the $\Upsilon(4S)$ almost exclusively decays into a $B\bar{B}$ pair.
Therefore, by exclusively reconstructing one $B$ meson and identifying the $K^{+}$ in the decay of the other $B$ meson, the missing mass technique can be used to reconstruct the $X(3872)$.
In the past, BaBar used a similar approach to perform this measurement with an data set of $231.8 \times 10^{6}$ $B\bar{B}$ pairs,
which resulted in an upper limit at 90$\%$ confidence level (C.L) of ${\cal B}(B^{+} \to X(3872) K^{+})  < 3.2 \times 10^{-4}$ \cite{Aubert:2005vi}.
The Belle collaboration measured the ${\cal B}(B^{+} \to X(3872) K^{+})  \times {\cal B}(X(3872) \to D^{0} \bar{D}^{0} \pi^{0})$ to be $(1.02 \pm 0.31^{0.21}_{0.29}) \times 10^{-4}$.
This value provides the lower limit of the ${\cal B}(B^{+} \to X(3872) K^{+})$ as ${\cal B}(X(3872) \to D^{0} \bar{D}^{0} \pi^{0})$ is smaller than unity.
In this paper, we present a measurement of  ${\cal B}(B^{+} \to X(3872) K^{+} )$ using the full data sample of the Belle experiment, along with
a simultaneous measurement of the various charmonium(-like) states ($X_{c\bar{c}}$) that appear in the missing mass spectrum.
We measure the branching fractions of nine states: $\eta_{c}$, $J/\psi$, $\chi_{c0}$, $\chi_{c1}$, $\eta_{c}(2S)$, $\psi(2S)$, $\psi(3770)$, $X(3872)$, and $X(3915)$.
In particular, the measurements of ${\cal B}(B^{+} \to \eta_{c} K^{+} )$ and ${\cal B}(B^{+} \to \eta_c(2S) K^{+} )$ are
important. Past measurements of these values~\cite{Aubert:2005vi} limited the precision of the determination of the absolute branching fractions of $\eta_{c}$ and $\eta_{c}(2S)$~\cite{PDG}.  
Also, this is the first limit for ${\cal B}(B^{+} \to X(3915) K^{+} )$.
Finally, we present the measurement of ${\cal B}(B^{+} \to  \bar{D}^{(\ast) 0} \pi^{+}) $, using a similar technique, where the $\bar{D}^{(\ast) 0}$
is reconstructed from the missing mass.
These are useful normalization modes for  
other $B$ meson decays such as $B^{+} \to \bar{D}^{(\ast) 0} K^{+}$ and $B^{+} \to  \bar{D}^{(\ast) 0} \pi^{+} \pi^{+} \pi^{-}$.

The remaining sections of the paper are organized as follows:
in Sec. \ref{section_data}, the Belle detector and the data samples used are described.
In Sec. \ref{section_selection}, an overview of the analysis method is provided.
In Sec. \ref{section_pi}, the analysis approach for the $B^{+} \to \bar{D}^{(\ast) 0} \pi^{+} $ decay is described.
In Sec. \ref{section_xcc}, the $B^{+} \to X_{c\bar{c}} K^{+} $ analysis is presented.
In Sec. \ref{section_sys}, the relevant systematic uncertainties are discussed.
Finally in Sec. \ref{section_conclusion}, the conclusion of this paper is presented.

\section{Data samples and the Belle detector}\label{section_data}
We use a data sample of $772\times10^{6}$ $B\bar{B}$ pairs recorded with the
Belle detector at the KEKB asymmetric-energy $e^{+}e^{-}$ collider \cite{KEKB}.
The Belle detector is a large-solid-angle magnetic
spectrometer that consists of a silicon vertex detector (SVD),
a 50-layer central drift chamber (CDC), an array of
aerogel threshold Cherenkov counters (ACC),  
a barrel-like arrangement of time-of-flight
scintillation counters (TOF), and an electromagnetic calorimeter
comprised of CsI(Tl) crystals (ECL) located inside 
a superconducting solenoid coil that provides a 1.5~T
magnetic field.  An iron flux-return located outside of
the coil is instrumented to detect $K_L^0$ mesons and to identify
muons.  The Belle detector is described in detail elsewhere~\cite{Belle}.

We use Monte-Carlo (MC) simulated events generated using {\tt EvtGen}~\cite{evtgen} and {\tt JETSET}~\cite{jetset} 
that include QED final-state radiation~\cite{photos}. The events are then processed by a detector simulation based on {\tt GEANT3}~\cite{GEANT}.
We produce signal MC events to obtain the reconstruction efficiency and the mass resolution for signal events.
We also use background MC samples to study the missing-mass distribution in the background process  
$\Upsilon(4S) \to B\bar{B}$ and $e^{+}e^{-} \to q\bar{q}$ ($q=u,d,s,c$ and $b$) with statistics six times that of data.

\section{Analysis overview}\label{section_selection}
In this analysis, we fully reconstruct one of the two charged $B$ mesons ($B_{\rm tag}$) via hadronic states
and require at least one charged kaon or pion candidate among the charged particles not used for the $B_{\rm tag}$ reconstruction.
The kaon or pion, coming from the other charged $B$ meson, $B_{\rm sig}$, is required to have a charge opposite that of $B_{\rm tag}$.
The $B_{\rm tag}$ is reconstructed in one of 1104 hadronic decays using a hierarchical hadronic
full reconstruction algorithm based on the NeuroBayes neural-network package \cite{Feindt:2011mr}.
The quality of a $B_{\rm tag}$ candidate is represented by a single NeuroBayes output-variable classifier 
($O_{\rm NB}$), which includes event-shape information to suppress continuum events.
We require $O_{\rm NB}$ to be greater than 0.01, which retains 90$\%$ of true $B_{\rm tag}$ candidates and 
rejects 70$\%$ of fake $B_{\rm tag}$ candidates. 
The beam constrained mass $M_{\rm bc} = \sqrt{E^{\ast 2}_{\rm beam}/c^{4} - |\vec{p}^{\ast 2}_{\rm tag}|/c^{2} }$,
where $E^{\ast}_{\rm beam}$ and $\vec{p}^{\ast}_{\rm tag}$ are the beam-energy and the reconstructed
$B_{\rm tag}$ three-momenta, respectively, in the center-of-mass frame, is required to be greater than 5.273 GeV/$c^{2}$.

\begin{figure*}[htbp]
  \begin{center}
    \includegraphics[scale=0.3]{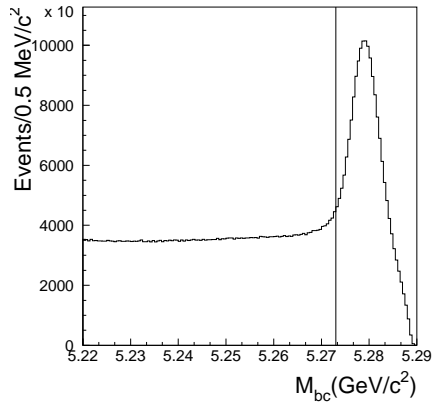}
    \caption{$M_{\rm bc}$ distribution for data after the requirement on the $O_{\rm NB}$.
              Vertical line shows the lower bound of the selection criteria.}
    \label{mbc}
  \end{center}
\end{figure*}

About 18$\%$ of the events contain multiple $B_{\rm tag}$ candidate that pass all the selection criteria.
In such an event, the $B_{\rm tag}$ with the greatest $O_{\rm NB}$ is retained.
The $B_{\rm tag}$ reconstruction efficiency is roughly 0.3$\%$.
Figure \ref{mbc} shows the $M_{\rm bc}$ distribution for data with the $O_{\rm NB}$ requirement applied. 
The selections of the charged kaon and pion daughters of $B_{\rm sig}$ are performed based on vertex information from 
the tracking system (SVD and CDC) and likelihood values $\mathcal{L}_{K}$ and $\mathcal{L}_{\pi}$
provided by the hadron identification system, ionization loss in the CDC,
the number of detected Cherenkov photons in the ACC, and the time-of-flight measured by the TOF~\cite{Nakano:2002jw}.
A charged track is required to have a point of closest approach to the interaction point that is within 5.0 cm along the $z$ axis and 
0.40 cm in the transverse ($r$-$\phi$) plane. The $z$ axis is opposite the positron beam direction.
A track is identified as a kaon (pion) if the likelihood ratio $\mathcal{L}(K:\pi)$ ($\mathcal{L}(\pi:K)$) is greater than 0.6.
The likelihood ratio is defined as $\mathcal{L}(i:j)=\mathcal{L}_{i}/(\mathcal{L}_{i}+\mathcal{L}_{j})$.
The efficiencies of hadron identification are about 90$\%$ for both pions and kaons.
The momentum-averaged probability to misidentify a pion (kaon) track as a kaon (pion) track is about 9$\%$ (10$\%$).
We identify the signal as a peak at the nominal $X_{c\bar{c}}$ or $\bar{D}^{\ast 0}$ mass in the distribution of missing mass:
\begin{equation}\label{equ_mm} 
M_{\rm miss(h)} = \sqrt{(p^{\ast}_{\rm e^{+}e^{-}}-p^{\ast}_{\rm tag} - p^{\ast}_{\rm h} )^{2}}/c,
\end{equation}
where $M_{\rm miss(h)}$ is the missing mass recoiling against the hadron $h$ ($\pi^{+}$ or $K^{+}$), and
$p^{\ast}_{e^{+}e^{-}}$, $p^{\ast}_{\rm tag}$, and $p^{\ast}_{h}$ are the four-momenta of the electron-positron initial state,
$B_{\rm tag}$, and $h$, respectively, in the center-of-mass frame. 
The probability to observe multiple kaon or pion candidates in the $M_{miss(h)}$ range of interest ($2.6$ GeV/$c^{2}$$<M_{miss(K^{+})}<$ 4.1 GeV/$c^{2}$
and $M_{miss(\pi^{+})}<$ 2.5 GeV/$c^{2}$) in an event is 2.8$\%$ and 0.3$\%$, respectively.
We do not apply a best-candidate selection if multiple candidates are found. 

The beam-energy resolution is a dominant contribution to the $M_{\rm bc}$ resolution,
and event-by-event fluctuations of $M_{\rm bc}$ from the nominal $B$ meson mass are directly correlated to the event-by-event fluctuation of the beam-energy.
We apply a correction to account for the event-by-event fluctuation of the beam-energy using a linear relation to $M_{\rm bc}$. 
The corrected beam-energy improves the missing mass resolution
by 8$\%$, 4$\%$, and 2$\%$ for $X(3872)$, $J/\psi$, and $D^{0}$, respectively.
The validity of the beam-energy correction is checked using high-statistics samples $B^{+} \to  D^{(\ast) 0} \pi^{+} $,
 $B^{+} \to D^{(\ast) 0} \pi^{+} \pi^{+} \pi^{-} $, and $B^{+} \to J/\psi K^{+}$ samples. We divide the samples into 
two sets with $M_{\rm bc}$ smaller or larger than the nominal $B^{+}$ mass~\cite{PDG}. 
The peak positions in the $M_{miss(h)}$ distribution for both data sets without the beam-energy correction 
are significantly different from their expected masses and are consistent within uncertainty after the correction.
We blinded the missing mass distribution in the range 3.3 GeV/$c^{2}$ $<$ $M_{miss(K^{+})}$ $<$ 4.0 GeV$/c^{2}$ 
until the analysis procedure was fixed. Branching fractions are obtained using the following equations:
\begin{eqnarray}\label{eq_branch}
\lefteqn{{\cal B }=\frac{N_{\rm sig}}{2N_{B^{\pm}} \epsilon}, }  \\
\lefteqn{N_{B^{\pm}}=N_{\Upsilon(4S)} {\cal B}(\Upsilon(4S) \to B^{+} B^{-}),} 
\end{eqnarray}
where $N_{\rm sig}$ is the signal yield obtained from the fit to the missing mass distribution,
$\epsilon$ is the reconstruction efficiency for $B_{\rm tag}$ and pion or kaon in $B_{\rm sig}$,
and $N_{\Upsilon(4S)}$ is the number of accumulated $\Upsilon(4S)$ events.
We use a value of $0.514$ for ${\cal B}(\Upsilon(4S) \to B^{+} B^{-})$ \cite{PDG}. 
The factor of two in Eq. (\ref{eq_branch}) originates from the inclusion of the charge-conjugate mode.

\section{Analysis of $B^{+} \to \bar{D}^{(\ast) 0} \pi^{+} $ decay}\label{section_pi}
\begin{figure*}[htbp]
  \begin{center}
    \includegraphics[scale=0.6]{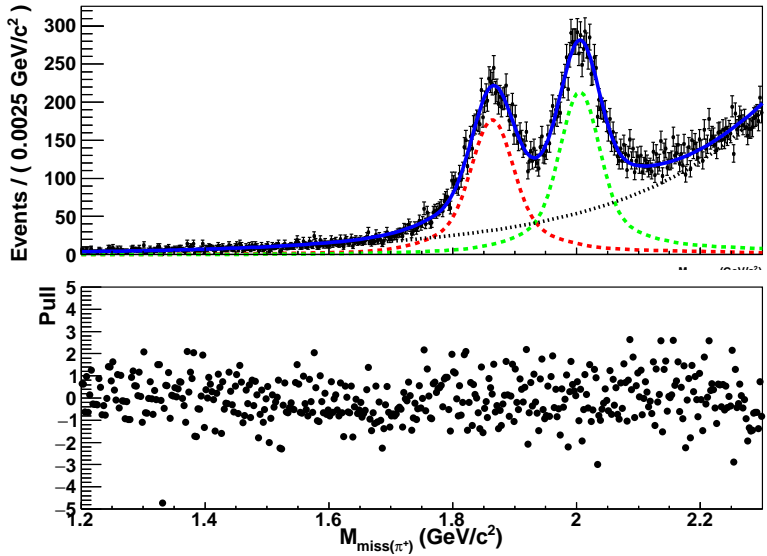}
    \caption{Observed $M_{miss(\pi^{+})}$ distribution. 
             Points with error bars represent data. 
             The solid, dashed, and dotted lines represent the total fit result, the contribution from $\bar{D}^{0}$ and $\bar{D}^{\ast 0}$,
             and the contribution from the background, respectively. }
    \label{mmpi}
  \end{center}
\end{figure*}

Figure \ref{mmpi} shows the observed $M_{miss(\pi^{+})}$ distribution, where clear peaks corresponding to
$\bar{D}^{0}$ and $\bar{D}^{\ast 0}$ are visible. In order to extract the signal $\bar{D}^{(\ast) 0}$ yields, a binned likelihood fit is performed.
The probability density function (PDF) for the signal peak is the sum of three Gaussian functions
based on a study of large simulated samples of signal decays. The mean value for one
Gaussian function is allowed to differ from that of the other two to accommodate for the tail in high-mass regions
resulting from $B_{\rm tag}$ decays with photons. The relative weights of the three Gaussian functions
are fixed to the values obtained from the signal MC. We introduce two parameters: the global offset of the mean $(\mu_{\rm data}-\mu_{\rm MC})$
and the global resolution scale factor $(\sigma_{\rm data}/\sigma_{\rm MC})$ to accommodate for a possible difference in the shape in 
the signal MC and data. The PDF for background events is represented by a second-order exponential:
$\exp (ax+bx^{2})$, where $a$ and $b$ are free parameters in the fit. The validity of using this function as a background PDF 
is confirmed by fitting to the background MC and sideband data, which is defined within the region $5.22$ GeV/$c^{2}<M_{\rm bc}<5.26$ GeV/$c^{2}$.
The fit returns a reasonable $\chi^{2}/ndf$, where $ndf$ is number of degree of freedom. $\chi^{2}$ is not improved by increasing the exponential order.
The mass range above 2.3 GeV$/c^{2}$ is not included in the fit to 
avoid contributions from excited $D$ mesons. 

Table \ref{summary_d} summarizes the branching fraction measurements for $B^{+} \to \pi^{+} \bar{D}^{(\ast) 0}$.
The values of $(\mu_{\rm data}-\mu_{\rm MC})$ and $(\sigma_{\rm data}/\sigma_{\rm MC})$ are found to be quite consistent at 0 MeV/$c^{2}$ and 1, respectively, indicating that the
signal MC describes the signal shape well. The measured branching fractions are consistent with the world averages \cite{PDG} within 1.1$\sigma$
taking into account the fact that almost all past measurements assumed ${\cal B}(\Upsilon(4S) \to B^{+} B^{-})=0.5$.  

\begin{center}
  \begin{table*}[htbp]
     \caption{Summary of the branching fraction measurements for $B^{+} \to  \bar{D}^{(\ast) 0 } \pi^{+}  $ decays. 
             The first uncertainties for the branching fractions are statistical and the second are systematic.}
     \begin{tabular}{c|ccccccc} \hline \hline
       Mode                    & $N_{\rm sig}$ &$(\mu_{\rm data}-\mu_{\rm MC})$ (MeV/$c^{2}$) &$(\sigma_{\rm data}/\sigma_{\rm MC})$ &$\epsilon (10^{-3})$ &  ${\cal B}$ ($10^{-3}$) & World average  for ${\cal B}$ ($10^{-3}$)~\cite{PDG}  \\ \hline 
       $B^{+}\to \pi^{+} \bar{D}^{0}$       & $8550\pm190$  &$-0.5\pm0.8$                                  &$0.994\pm0.025$                       &2.48$\pm0.02$        &  $4.34\pm0.10\pm0.25$   & $4.80\pm0.15$           \\
       $B^{+}\to \pi^{+} \bar{D}^{\ast 0}$  & $9980\pm250$  &$-0.8\pm0.8$                                  &$1.035\pm0.029$                       &2.61$\pm0.02$        &  $4.82\pm0.12\pm0.35$   & $5.18\pm0.26$                         \\ \hline \hline
    \end{tabular}    
    \label{summary_d}
  \end{table*}
\end{center}

\section{Analysis of $B^{+} \to X_{c\bar{c}} K^{+} $ decay} \label{section_xcc}
\begin{figure*}[htbp]
  \begin{center}
    \includegraphics[scale=0.7]{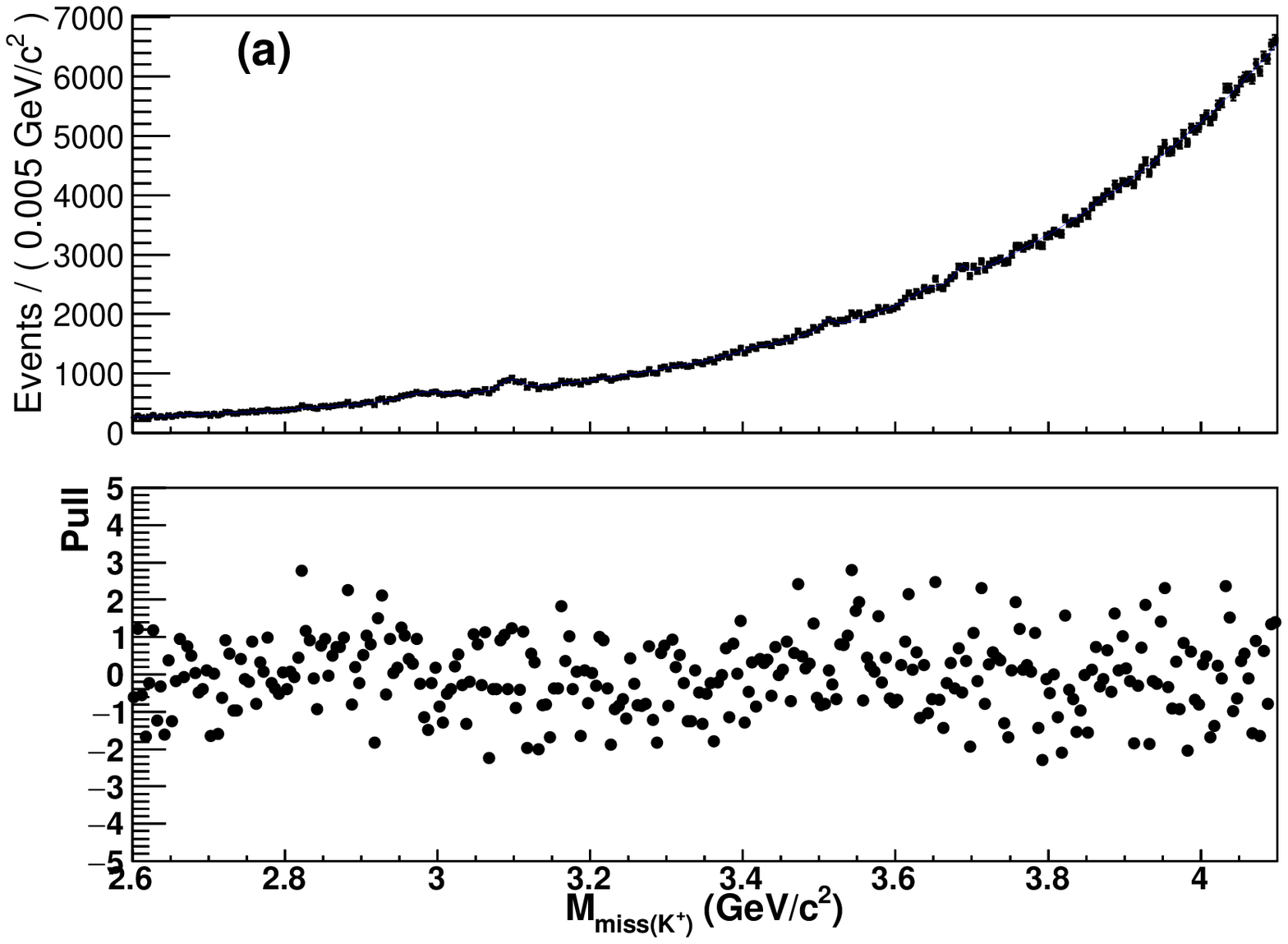}
    \includegraphics[scale=0.35]{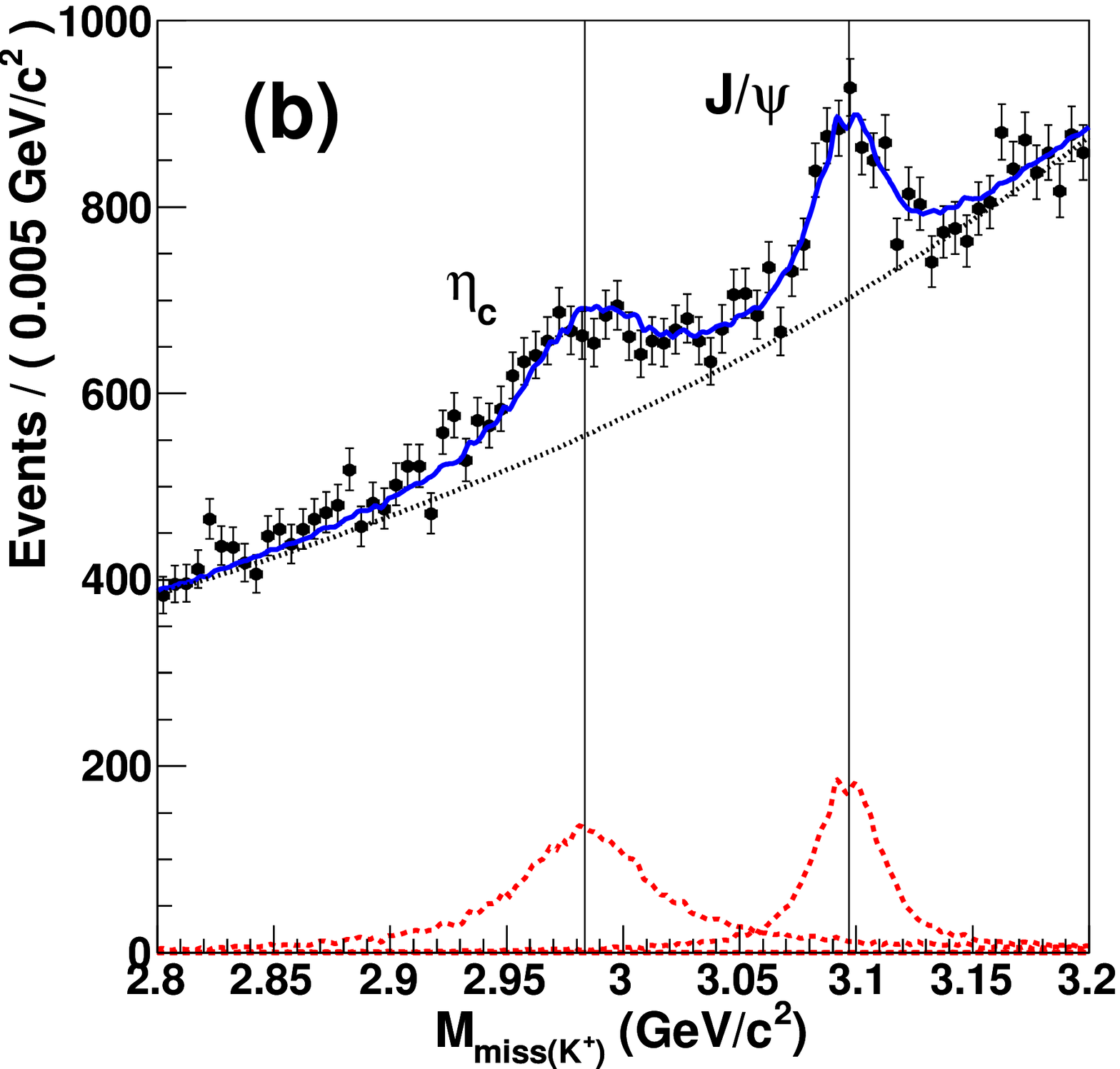}
    \includegraphics[scale=0.35]{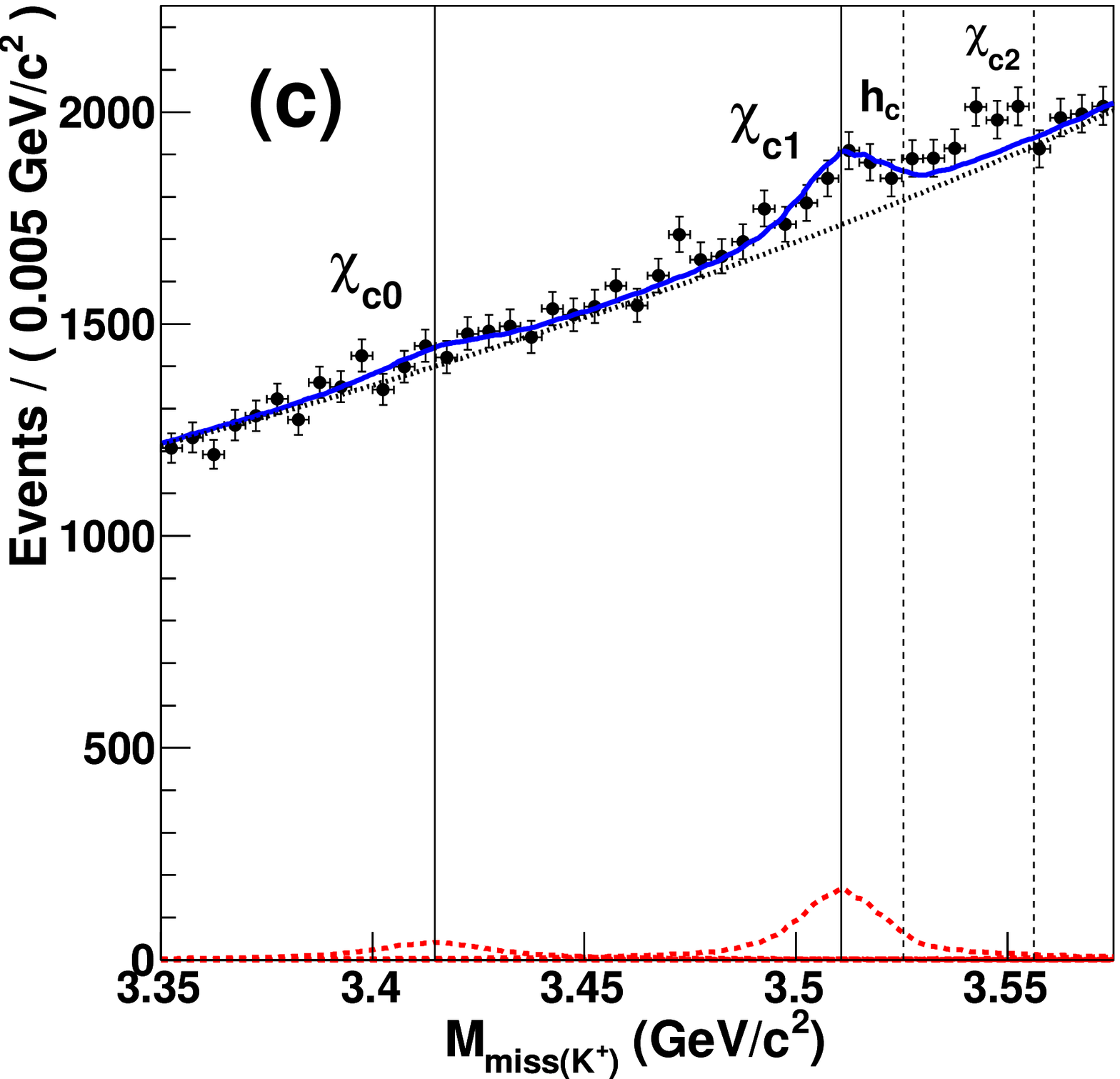}
    \includegraphics[scale=0.35]{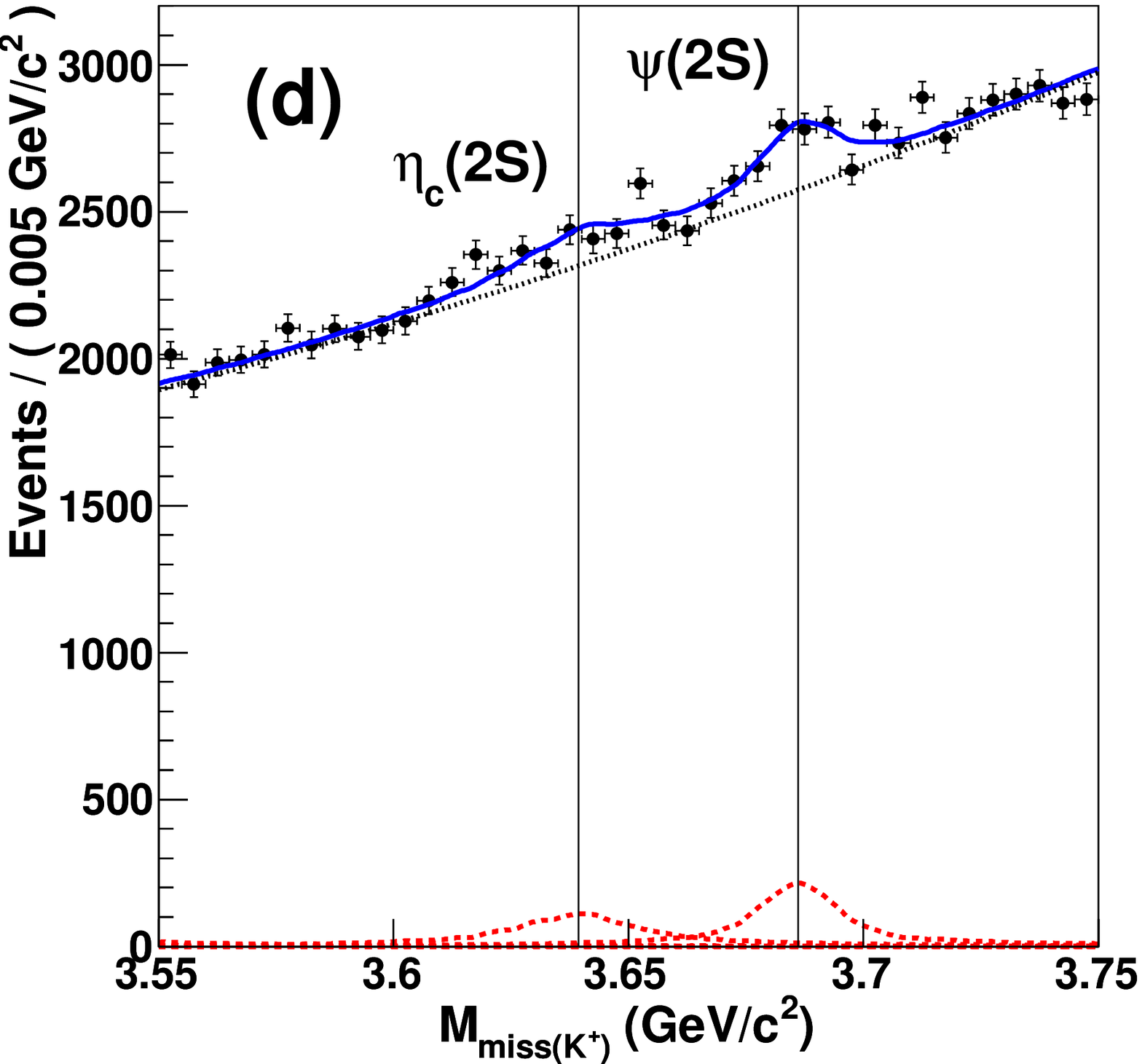}
    \includegraphics[scale=0.35]{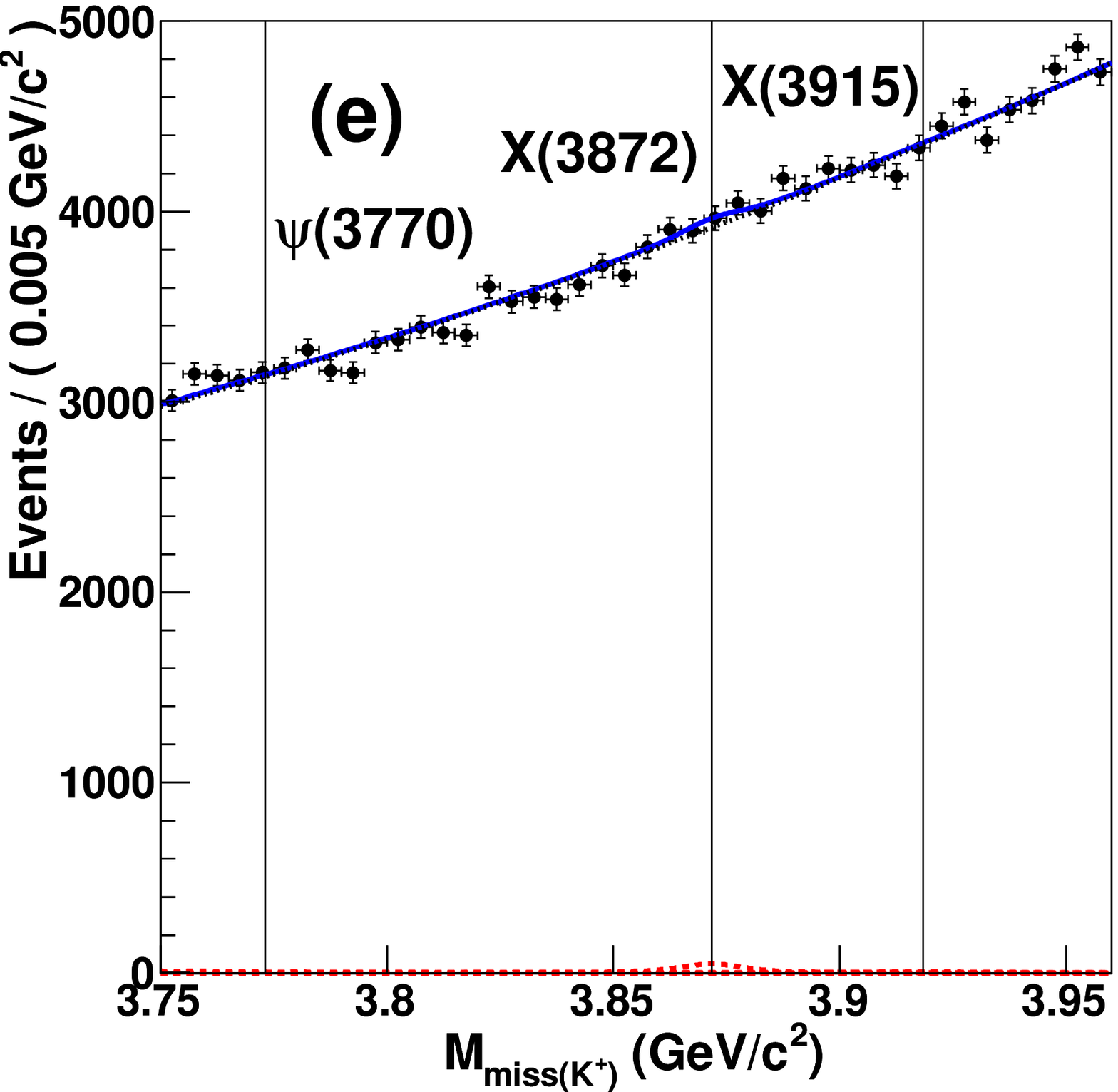}
    \caption{Observed $M_{\rm miss(K^{+})}$ distributions:(a) shows the full $M_{\rm miss(K^{+})}$ region, with pull distribution,
            and (b), (c), (d), and (e) are for zoomed plots for specific $X_{c\bar{c}}$. Points with error bars represent data. 
            Vertical solid lines show the nominal mass of $X_{c\bar{c}}$ included in the fit.
            Vertical dashed lines show the ones not included in the fit.
            Solid line represents the total fit result. Dashed and dotted lines are $X_{c\bar{c}}$ contributions and background contributions, respectively.
            }
    \label{mmk}
  \end{center}
\end{figure*}

Figure \ref{mmk} shows the observed and fitted $M_{\rm miss(K^{+})}$ distributions.
We again perform a binned likelihood fit to extract the signal $X_{c\bar{c}}$ yields.
In the analysis of the high-statistics sample of $B^{+} \to  \bar{D}^{(\ast) 0} \pi^{+} $, we confirm that the
signal shape is consistent between data and MC. 
Therefore, we fix the signal PDF to be the histogram PDF from signal MC generated
with the mass and natural width of the $X_{c\bar{c}}$ states fixed to the world averages~\cite{PDG}. We consider nine $X_{c\bar{c}}$ in the fit:
$\eta_{c}$, $J/\psi$, $\chi_{c0}$, $\chi_{c1}$, $\eta_{c}(2S)$, $\psi(2S)$, $\psi(3770)$, $X(3872)$, and $X(3915)$.
We do not include $h_{c}$ and $\chi_{c2}$ because their branching fractions are measured to be very small~\cite{PDG}:
${\cal B}(B^{+} \to K^{+} h_{c})<3.8\times 10^{-5}$ at 90$\%$ C.L. and ${\cal B}(B^{+} \to K^{+} \chi_{c2})=(1.1 \pm 0.4 )\times 10^{-5}$.
The background PDF is a second-order exponential, as for $B^{+} \to \pi^{+} \bar{D}^{(\ast)0}$, and is again validated with background MC
and the data sideband.  
The statistical significance of each $X_{c\bar{c}}$ state is determined from the log-likelihood ratio 
$-2\ln{(\mathcal{L}_{0}/\mathcal{L})}$, where $\mathcal{L}_{0}$ ($\mathcal{L}$) is the likelihood for the fit without (with) the signal component.
When we evaluate the significance for a $X_{c\bar{c}}$ state, the other $X_{c\bar{c}}$ states are included in the fit.
The branching fractions are determined using Eq. (\ref{eq_branch}). 
For $\chi_{c0}$, $\psi(3770)$, and $X(3872)$, the significances are smaller than three standard deviations ($\sigma$). 
We also set 90$\%$ C.L. upper limit for branching fractions to these states using the $CLs$ technique~\cite{cls}.

The results are summarized in Table \ref{summary_xcc}.
The upper limit for ${\cal B}(B^{+} \to X(3872) K^{+} )$ is the most stringent to date.
The upper limit for the ${\cal B}(B^{+} \to X(3915) K^{+})$ is determined for the first time.
The measurements for ${\cal B}(B^{+} \to \eta_{c}  K^{+} )$ and ${\cal B}(B^{+} \to \eta_{c}(2S) K^{+} )$ are the most precise to date.
In particular, this is the first significant measurement of ${\cal B}(B^{+} \to \eta_{c}(2S) K^{+})$.
For ${\cal B}(B^{+} \to \psi(3770)  K^{+} )$, we do not see a significant signal and the measured value is smaller than world average by 2.7$\sigma$.
For the other measurements, the values are consistent with world averages within 1.7$\sigma$.

In Figure \ref{mmk} (c), we see an enhancement near 3545 MeV/$c^{2}$, where no known charmonium state exists.
We attempt to fit this by including an additional contribution using signal MC PDF with a mass of 3545 MeV/$c^{2}$ 
and a natural width of 0 MeV. An offset for the peak position is introduced as a free parameter.
The signal yield is $738\pm189$ events at the peak position of $3544.2\pm2.8$ MeV/$c^{2}$, which is  
4.3$\sigma$ lower than the mass of the $\chi_{c2}$. The $-2\ln{(\mathcal{L}_{0}/\mathcal{L})}$ value is 16.5.
Since the signal region is very wide compared to the experimental resolution, we estimate the probability to observe such an enhancement in a single experiment.
We perform one million pseudo-experiments in which background events are generated with the same shape and yield as data.
Multiple fits, each including signal with a natural width of zero and a mass value incremented by 1 MeV/c$^{2}$
across the fit range for successive fits, are performed and the highest $-2\ln{(\mathcal{L}_{0}/\mathcal{L})}$ in one pseudo-experiment is retained.
The probability to observe an enhancement with $-2\ln{(\mathcal{L}_{0}/\mathcal{L})}$  
greater than 16.5 is 0.43$\%$, which corresponds to a global significance of 2.8$\sigma$. 
We therefore conclude that the enhancement is not significant.

\begin{center}
  \begin{table*}[htbp]
    \caption{Summary of the branching fraction measurements for $B^{+} \to X_{c\bar{c}} K^{+} $ decay.
             For the branching fractions, the first uncertainties are statistical and the second are systematic.
             Values in brackets for ${\cal B}$ represent the 90$\%$ C.L. upper limits. }
     \begin{tabular}{c|ccccc} \hline \hline
       Mode                 &Yield        &Significance ($\sigma$) & $\epsilon (10^{-3})$     &${\cal B}$ ($10^{-4}$)      & World average for ${\cal B}$ ($10^{-4}$)~\cite{PDG}\\ \hline
       $\eta_{c}$           &$2590\pm180$ &14.2                    & $2.73\pm0.02$            &$12.0\pm0.8\pm0.7$          & $9.6\pm1.1$\\
       $J/\psi$             &$1860\pm140$ &13.7                    & $2.65\pm0.02$            &$8.9\pm0.6\pm0.5$           & $10.26\pm0.031$\\
       $\chi_{c0}$          &$430\pm190$  &2.2                     & $2.67\pm0.02$            &$2.0\pm0.9\pm0.1$  $(<3.3)$ & $1.50^{+0.15}_{-0.14}$\\
       $\chi_{c1}$          &$1230\pm180$ &6.8                     & $2.68\pm0.02$            &$5.8\pm0.9\pm0.5$           & $4.79\pm0.23$\\
       $\eta_{c}(2S)$       &$1050\pm240$ &4.1                     & $2.77\pm0.02$            &$4.8\pm1.1\pm0.3$           & $3.4\pm1.8$\\
       $\psi(2S)$           &$1410\pm210$ &6.6                     & $2.79\pm0.02$            &$6.4\pm1.0\pm0.4$           & $6.26\pm0.24$\\
       $\psi(3770)$         &$-40\pm310$  &-                       & $2.76\pm0.02$            &$-0.2\pm1.4\pm0.0$ $(<2.3)$ & $4.9\pm1.3$\\
       $X(3872)$            &$260\pm230$  &1.1                     & $2.79\pm0.01$            &$1.2\pm1.1\pm0.1$  $(<2.6)$ & $(<3.2)$ \\
       $X(3915)$            &$80\pm350$   &0.3                     & $2.79\pm0.01$            &$0.4\pm1.6\pm0.0$  $(<2.8)$ &  -  \\ \hline \hline 
    \end{tabular}    
    \label{summary_xcc}
  \end{table*}
\end{center}

\section{Systematic uncertainty} \label{section_sys}
A summary of the systematic uncertainties for each $X_{c\bar{c}}$ and $\bar{D}^{(\ast) 0}$ state is provided in Table \ref{summary_sys}.
We consider the following systematic uncertainties for the branching fraction measurements.
The systematic uncertainty for the efficiency of the charged hadron identification is estimated from the yield of 
$D^{\ast +} \to D^{0}\pi^{+}$, $D^{0}\to K^{-}\pi^{+}$ with and without the hadron identification requirements.
We apply a correction factor to the particle identification efficiencies based on the ratio of the efficiencies found in the MC and data samples.
The uncertainty of the correction factor is treated as the systematic uncertainty.
The systematic uncertainty due to the charged track reconstruction efficiency is estimated
using the decay chain $D^{\ast +}\to \pi^{+}D^{0}$, $D^{0}\to \pi^{+}\pi^{-}K^{0}_{S}$, and $K^0_{S}\to\pi^{+}\pi^{-}$
where $K^{0}_{S} \to \pi^{+} \pi^{-}$ is either partially or fully reconstructed. The ratio between the yields
of partially and fully reconstructed signals are compared between data and MC; the difference of 0.35$\%$ per track is taken as the 
systematic uncertainty. 
The systematic uncertainty from the reconstruction efficiency of $B_{\rm tag}$ is estimated using a hadronic-tag analysis in which $B_{\rm sig}$
decays to $D^{(\ast)} \ell \nu$, where $l$ is electron or muon~\cite{Sibidanov:2013rkk}.
The yield for this signal is compared between data and MC, and the difference is implemented as 
a correction factor for each $B_{\rm tag}$ decay mode. The averaged correction factor used in this analysis is
0.76 independent of $X_{c\bar{c}}$. The signal and background efficiencies are multiplied by this factor.
The main origin of the correction factor is understood to result from the fact that branching fractions
for some of the $B_{\rm tag}$ decays in the MC generation are outdated and inconsistent withe the most recent measurements.
Furthermore, this correction factor is also determined in the $B^{-} \to \tau^{-} \nu$ analysis independently using sideband region
of extra ECL energy \cite{Adachi:2012mm}. The correction factor is found to be consistent
with that obtained from $D^{(\ast)} \ell \nu$, which indicates that it is generally independent of the $B_{sig}$ decay mode.
The uncertainty of this correction factor, 4.6$\%$, is regarded as the systematic uncertainty.
Note that the correction for lepton identification efficiency and associated systematic uncertainty were not
taken into account in Ref.~\cite{Sibidanov:2013rkk}. These are implemented in this analysis. 
The world average of ${\cal B}(\Upsilon(4S) \to B^{+} B^{-})$ is ($51.4 \pm 0.6 $)$\%$ \cite{PDG}, which corresponds to 
a systematic uncertainty of 1.2$\%$.
The systematic uncertainty for the $N_{\Upsilon(4S)}$ is assigned as 1.4$\%$.
The systematic uncertainty due to the mass and width of each state is estimated by  
performing fits while varying the mass and width by the world-average uncertainties~\cite{PDG}.
The systematic uncertainty on the fitter bias is estimated by performing pseudo-experiments.
We generate pseudo-data from the background and signal shapes determined from the background MC and the signal MC, respectively.
The background yields are taken from the sideband of data defined as $M_{\rm miss(K^{+})}$ $<$ 3.3 GeV/$c^{2}$ or  $M_{\rm miss(K^{+})}$ $>$ 4.0 GeV$/c^{2}$,
and signal yields are determined from the world averages of the branching fractions and the reconstruction efficiencies.
We perform a binned likelihood fit to extract the signal yields in each pseudo-experiment.
The difference between the mean of the extracted signal yields and the input mass value is taken as the systematic uncertainty. 
The systematic uncertainty due to the $B^{+} \to X_{c\bar{c}}K^{+}$ signal PDF arises from uncertainties of $\mu_{\rm data} - \mu_{\rm MC}$
and $\sigma_{\rm data}/\sigma_{\rm MC}$ in the $B^{+} \to  D^{0 (\ast)} \pi^{+} $ fit. This is evaluated by performing a fit with the
PDF shape parameters changed within their uncertainties after averaging the result for $\bar{D}^{0}$ and $\bar{D}^{\ast 0}$.
 All the charmonium states are changed simultaneously in this case.
The systematic uncertainty arising from the finite statistics of the signal MC is estimated by repeatedly modifying the histogram-PDF bin contents
within the Poisson uncertainty and then refitting. 
The root mean square of the extracted signal yield distribution is regarded as the systematic uncertainty.  

The reconstruction efficiency of the $B_{tag}$ may depend on the decay of $X_{c\bar{c}}$ states.
As our knowledge of the decay modes of each $X_{c\bar{c}}$ is limited, a systematic uncertainty is assigned in the following way.
For each $X_{c\bar{c}}$ except $\psi(3770)$, the sum of the known branching fractions is not equal to 100$\%$.
In the default estimation of the branching fraction, unknown decay modes are filled with decays into
$u\bar{u}, d\bar{d}$, and $s\bar{s}$, that hadronize 
via PYTHIA. The systematic uncertainty is estimated by eliminating the PYTHIA-generated decay and taking 
the difference with the nominal efficiency. For $X(3872)$ and $X(3915)$, the PYTHIA decay is not implemented by default.
Therefore, the systematic uncertainty is estimated by implementing the PYTHIA decay with a branching fraction of 50$\%$.
The systematic uncertainty on the background assumption is estimated by performing a fit after changing the 
order of the exponential background's polynomial exponent from quadratic to cubic.
We perform a fit including $h_{c}$ and $\chi_{c2}$ with yields fixed to their world averages and upper limit
for the branching fractions, respectively.
The difference of the yields from the default fit is regarded as a systematic uncertainty.
The statistical uncertainty of the  signal reconstruction efficiency due to signal MC statistics is regarded as systematic uncertainty.

\begin{center}
  \begin{table*}[htbp]
    \caption{Summary of the systematic uncertainties ($\%$).}
     \begin{tabular}{c|ccccccccccc} \hline \hline
       Source                                             & $\eta_{c}$ & $J/\psi$  & $\chi_{c0}$ & $\chi_{c1}$  & $\eta_{c}(2S)$ & $\psi(2S)$  & $\psi(3770)$& $X(3872)$& $X(3915)$    &  $\bar{D}^{0}$  & $\bar{D}^{\ast 0}$  \\ \hline
       PID                                                &0.8          &0.8        &0.7          & 0.7          & 0.8            & 0.7         &  0.8        &  0.9       &   0.9      & 0.9             &    0.9        \\
       Tracking                                           &0.35         &0.35       &0.35         & 0.35         & 0.35           & 0.35        &  0.35       &  0.35      &   0.35     & 0.35            &    0.35       \\
       $B_{\rm tag}$                                      &4.6          &4.6        &4.6          & 4.6          & 4.6            & 4.6         &  4.6        &  4.6       &   4.6      & 4.6             &    4.6        \\
       ${\cal B}(\Upsilon(4S))$                           &1.2          &1.2        &1.2          & 1.2          & 1.2            & 1.2         &  1.2        &  1.2       &   1.2      & 1.2             &    1.2        \\
       $N_{\Upsilon (4S)}$                                &1.4          &1.4        &1.4          & 1.4          & 1.4            & 1.4         &  1.4        &  1.4       &   1.4      & 1.4             &    1.4        \\
       M $\&$ $\Gamma$ (PDG)                              &1.5          &0.4        &1.6          & 0.2          & 0.8            & 0.4         &  0.4        &  0.1       &   5.7      & -               &    -          \\
       Fit bias                                           &0.0          &0.2        &1.9          & 0.6          & 0.4            & 1.0         &  0.7        &  0.6       &   1.6      & 0.2             &    0.2        \\
 $\mu_{\rm data} - \mu_{\rm MC}$ and  $\sigma_{\rm data}/\sigma_{\rm MC}$ &0.3      &1.5          &0.8          & 3.6          & 1.0            & 3.7         &  1.7        &  3.4       &   2.0      & -               &    -          \\
       Histogram PDF                                       &0.7          &0.5       &3.2          & 1.5          & 2.4            & 1.5         &  2.5        &  0.9       &   2.4      & -               &    -          \\
       Decay simulation                                   &2.1          &3.0        &1.1          & 2.0          & 1.3            & 1.0         &  0.0        &  0.9       &   0.5      & -               &    -          \\
       Background parametrization                        &0.6          &0.3         &1.7          & 0.6          & 3.8            & 0.2         &  2.0        &  1.8       &   3.5      & 2.8             &    5.0        \\
       $\chi_{c2}$/$h_{c}$                                &0.0          &0.0        &1.3          & 3.9          & 0.7            & 0.6         &  1.1        &  0.4       &   2.5      & -               &    -          \\ 
       Signal MC statistics                               &0.8          &0.8        &0.8          & 0.8          & 0.8            & 0.8         &  0.8        &  0.4       &   0.8      & 0.8             &    0.8        \\ \hline \hline
       Total                                              &5.8          &6.1        &7.0          & 7.8          & 7.1            & 6.7         &  6.4        &  6.5       &   9.5      & 5.8             &    7.2        \\
    \end{tabular}    
    \label{summary_sys}
  \end{table*}
\end{center}

\section{Conclusion} \label{section_conclusion}
We present the measurement of the absolute branching fractions for $B^{+} \to X_{c\bar{c}} K^{+} $, where $X_{c\bar{c}}$ denotes 
$\eta_{c}$, $J/\psi$, $\chi_{c0}$, $\chi_{c1}$, $\eta_{c}(2S)$, $\psi(2S)$, $\psi(3770)$, $X(3872)$, and $X(3915)$,
and also $B^{+} \to \pi^{+} D^{(\ast) 0}$. 
We do not observe a significant signal for $X(3872)$ and set an 90$\%$ C.L. upper limit of ${\cal B}(B^{+} \to X(3872) K^{+} )<2.6 \times 10^{-4}$, which is more stringent 
than the one determined by BaBar~\cite{Aubert:2005vi} ($3.2\times10^{-4}$).
The lower limit of ${\cal B} (X(3872) \to f)$ is based on BaBar's measurement. 
Our result improves these lower limits. We set the 90$\%$ C.L. upper limit of ${\cal B}(B^{+} \to X(3915) K^{+} )<2.8 \times 10^{-4}$ for the first time.\\
$\ $We measure ${\cal B}(B^{+} \to \eta_{c} K^{+} )=(12.0\pm0.8\pm0.7) \times 10^{-4}$ and ${\cal B}(B^{+} \to \eta_{c}(2S) K^{+} )=(4.8\pm1.1\pm0.3) \times 10^{-4}$,
which are the most accurate measurements to date. In particular, this is the first significant measurement for ${\cal B}(B^{+} \to \eta_{c}(2S) K^{+})$.
The current world average of ${\cal B}(\eta_{c}(2S) \to K\bar{K}\pi)$ is $(1.9\pm 0.4 \pm 1.1) \%$~\cite{PDG}, where the second uncertainty is dominated by  
the measurement of ${\cal B}(B^{+} \to \eta_{c}(2S) K^{+})=(3.4\pm1.8)\times 10^{-4}$ by BaBar~\cite{Aubert:2005vi}.
Our measurement significantly improves the precision of ${\cal B}(\eta_{c}(2S) \to K\bar{K}\pi)$. 
In addition, this measurement can contribute to many other decays involving the $\eta_{c}(2S)$ such as $\psi(2S) \to \gamma \eta_{c}(2S)$ by BESIII~\cite{Ablikim:2012sf}
and $\eta_{c}(2S) \to p\bar{p}$ by LHCb~\cite{Aaij:2016kxn}. Finally, we measure ${\cal B}(B^{+} \to \bar{D}^{0}  \pi^{+} )=(4.34\pm0.10\pm0.25) \times 10^{-3}$ and ${\cal B}(B^{+} \to \bar{D}^{\ast 0} \pi^{+} )=(4.82\pm0.12\pm0.35) \times 10^{-3}$,
which are consistent with the world averages \cite{PDG}. The latter is the most precise measurement from a single experiment.

\acknowledgments
We thank the KEKB group for the excellent operation of the
accelerator; the KEK cryogenics group for the efficient
operation of the solenoid; and the KEK computer group,
the National Institute of Informatics, and the 
PNNL/EMSL computing group for valuable computing
and SINET5 network support.  We acknowledge support from
the Ministry of Education, Culture, Sports, Science, and
Technology (MEXT) of Japan, the Japan Society for the 
Promotion of Science (JSPS), and the Tau-Lepton Physics 
Research Center of Nagoya University; 
the Australian Research Council;
Austrian Science Fund under Grant No.~P 26794-N20;
the National Natural Science Foundation of China under Contracts 
No.~10575109, No.~10775142, No.~10875115, No.~11175187, No.~11475187, 
No.~11521505 and No.~11575017;
the Chinese Academy of Science Center for Excellence in Particle Physics; 
the Ministry of Education, Youth and Sports of the Czech
Republic under Contract No.~LTT17020;
the Carl Zeiss Foundation, the Deutsche Forschungsgemeinschaft, the
Excellence Cluster Universe, and the VolkswagenStiftung;
the Department of Science and Technology of India; 
the Istituto Nazionale di Fisica Nucleare of Italy; 
the WCU program of the Ministry of Education, National Research Foundation (NRF)
of Korea Grants No.~2011-0029457, No.~2012-0008143,
No.~2014R1A2A2A01005286,
No.~2014R1A2A2A01002734, No.~2015R1A2A2A01003280,
No.~2015H1A2A1033649, No.~2016R1D1A1B01010135, No.~2016K1A3A7A09005603, No.~2016K1A3A7A09005604, No.~2016R1D1A1B02012900,
No.~2016K1A3A7A09005606, No.~NRF-2013K1A3A7A06056592;
the Brain Korea 21-Plus program, Radiation Science Research Institute, Foreign Large-size Research Facility Application Supporting project and the Global Science Experimental Data Hub Center of the Korea Institute of Science and Technology Information;
the Polish Ministry of Science and Higher Education and 
the National Science Center;
the Ministry of Education and Science of the Russian Federation and
the Russian Foundation for Basic Research;
the Slovenian Research Agency;
Ikerbasque, Basque Foundation for Science and
MINECO (Juan de la Cierva), Spain;
the Swiss National Science Foundation; 
the Ministry of Education and the Ministry of Science and Technology of Taiwan;
and the U.S.\ Department of Energy and the National Science Foundation.
This work is supported by a Grant-in-Aid for Scientific Research (S) ``Probing New Physics with Tau-Lepton'' (No.26220706),
Grant-in-Aid for Scientific Research on Innovative Areas ``Elucidation of New Hadrons with a Variety of Flavors'',
Grant-in-Aid from MEXT for Science Research in a Priority Area (``New Development of Flavor Physics'')
and from JSPS for Creative Scientific Research (``Evolution of Tau-lepton Physics'').

\end{document}